\documentclass[twocolumn]{aastex631}
\usepackage{CJK}
\usepackage{multirow}

\shorttitle{Disk occultation events}
\shortauthors{Zhu et al.}

\begin{document}
\begin{CJK*}{UTF8}{gbsn}

\title{Two Candidate KH 15D-like Systems from the Zwicky Transient Facility}

\author[0000-0003-4027-4711]{Wei Zhu (祝伟)}
\affiliation{Department of Astronomy, Tsinghua University, Beijing 100084, China}

\author{Klaus~Bernhard}
\affiliation{Bundesdeutsche Arbeitsgemeinschaft fuer Veraenderliche Sterne e.V. (BAV), Berlin, Germany}

\author[0000-0002-8958-0683]{Fei Dai (戴飞)}
\affiliation{Division of Geological and Planetary Sciences, California Institute of Technology, Pasadena, CA 91125, USA}

\author[0000-0001-8060-1321]{Min Fang (房敏)}
\affiliation{Purple Mountain Observatory, Chinese Academy of Sciences, 10 Yuanhua Road, Nanjing 210023, China}

\author[0000-0002-9849-5886]{J.~J.~Zanazzi}
\affiliation{Canadian Institute for Theoretical Astrophysics, University of Toronto, 60 St George Street, Toronto, ON M5S 3H8, Canada}

\author[0000-0001-6000-3463]{Weicheng Zang (臧伟呈)}
\affiliation{Department of Astronomy, Tsinghua University, Beijing 100084, China}

\author[0000-0002-1027-0990]{Subo Dong (东苏勃)}
\affiliation{Kavli Institute for Astronomy and Astrophysics, Peking University, Beijing 100871, China}

\author{Franz-Josef~Hambsch}
\affiliation{Bundesdeutsche Arbeitsgemeinschaft fuer Veraenderliche Sterne e.V. (BAV), Berlin, Germany}
\affiliation{Vereniging Voor Sterrenkunde (VVS), Brugge, BE-8000, Belgium}

\author[0000-0002-4503-9705]{Tianjun Gan (干天君)}
\affiliation{Department of Astronomy, Tsinghua University, Beijing 100084, China}

\author{Zexuan Wu (吴泽炫)}
\affiliation{Kavli Institute for Astronomy and Astrophysics, Peking University, Beijing 100871, China}
\affiliation{Department of Astronomy, School of Physics, Peking University, Beijing 100871, China}

\author[0000-0001-7739-9767]{Michael Poon}
\affiliation{Department of Astronomy and Astrophysics, University of Toronto, Toronto, ON M5S 3H4, Canada}

\begin{abstract}
KH 15D contains a circumbinary disk that is tilted relative to the orbital plane of the central binary. The precession of the disk and the orbital motion of the binary together produce rich phenomena in the photometric light curve. In this work, we present the discovery and preliminary analysis of two objects that resemble the key features of KH 15D from the Zwicky Transient Facility. These new objects, Bernhard-1 and Bernhard-2, show large-amplitude ($>1.5\,$mag), long-duration (more than tens of days), and periodic dimming events. A one-sided screen model is developed to model the photometric behaviour of these objects, the physical interpretation of which is a tilted, warped circumbinary disk occulting the inner binary. Changes in the object light curves suggest potential precession periods over timescales longer than 10 years. Additional photometric and spectroscopic observations are encouraged to better understand the nature of these interesting systems.
\end{abstract}



\section{Introduction} \label{sec:intro}

Kearns-Herbst 15D \citep[KH 15D,][]{Kearns1998} represents a rare class of circumbinary disk system. Photometric observations that went all the way back to the 1950s showed a complex light curve behaviour \citep{Johnson2005, Maffei2005, 2005Hamilton, Capelo(2012), Aronow2018, GarciaSoto2020}, characterized by the periodic (with period of 48\,days) dippings and the decades-long dimming and re-brightening. Together with the spectroscopic observations of the central object \citep{Johnson2004}, studies have shown that the circumbinary disk is largely tilted relative to the central, highly eccentric binary \citep{Chiang2004, Winn:2006, SilviaAgol(2008), GarciaSoto2020, Poon:2021}. Observations and more detailed studies of systems like KH 15D can provide useful constraints and insights into the evolution and dynamics of circumbinary disks (see \citealt{Poon:2021} and references therein).

Motivated by this, one of our co-authors, K.\ Bernhard, as an amateur astronomer, performed his search to identify similar systems in the ongoing all-sky variability surveys. Specifically, his search was focused on the variable star catalog of \citet{Chen:2020}, which was based on data collected by the Zwicky Transient Facility \citep[ZTF,][]{Bellm:2019, Masci:2019}. Several other authors of the present work were notified by Bernhard about the potentially KH 15D-like candidates later. This eventually led to the further analysis and observations of the identified systems, as will be presented in the rest of this work. 

\section{Candidate Search}

The primary feature of the photometric light curve of KH 15D is its deep ($\sim4\,$mag in $I$) and long ($\sim50\%$ of the binary period) occultation event on an otherwise photometrically relatively quiet star.
\footnote{As a weak-lined T Tauri star, KH 15D is variable at the level of $\sim0.1\,$mag due to the spots and stellar rotation \citep{2005Hamilton}. }
In addition, the light curve of KH 15D shows gradual changes in the baseline flux over a timescale of decades, due to the precession of the warped circumbinary disk. With only relatively short (i.e., a few years) time baseline, this latter feature is not expected to seen in the ZTF data.

The search starts from the ZTF variable star catalog of \citet{Chen:2020}, which contains about 780,000 periodic variables with classifications and another about 1,000,000 suspected variables with no classifications. 
Our proto-type, KH 15D, belongs to the second catalog, probably because it is a rare type of variables and could not be classified as any of the variable types of \citet{Chen:2020}. Motivated by this, we focused the search for more KH 15D-like objects on the suspected variable catalog of \citet{Chen:2020}.
\footnote{For completeness, we also checked the variable catalog with classifications (Table 2 of \citealt{Chen:2020}. There are only four sources with large enough ($>2\,$mag) variations, all with short ($<4\,$days) periods.}

Our final search follows closely the original procedure of K.\ Bernhard. First, variables with small photometric amplitudes (defined as $<1.5\,$mag in both $g$ and $r$ bands) or short periods (i.e., $<10\,$days in either $g$ or $r$) are excluded. This substantially reduces the sample to 1041. Next, we perform the box least squares (BLS, \citealt{Kovacs:2002, Hartman:2016}) analysis to all survival variables, to search for transit-like signals that closely resemble the occultation light curve of KH 15D. In the BLS analysis, we restrict the duration to be between 1\,day and 90\% of the searched period. The lower bound is effective in excluding the majority of eclipsing binaries that escaped from the classification of \citet{Chen:2020}, whereas the upper bound is useful in excluding variables with periodic outbursts (e.g., U Gem-type variables). Finally, all BLS analysis results are visually inspected to identify the probable candidates. The inspection primarily checks the unfolded and folded light curves and the level of variations outside of the best-fit ``transit duration'' window. 

The systematic search reveals no more promising candidates other than KH 15D and the three candidates that were originally identified by K.\ Bernhard. 
After a closer look into the light curves, we accept the two most promising candidates for further analysis. These are assigned the names \emph{Bernhard-1} and \emph{Bernhard-2}. The third candidate, ZTFJ070412.91-112403.2, is rejected because the photometric scatterings in its out-of-occultation light curve are comparable ($\sim1$ mag vs.\ $1.5$ mag in $r$) to the depth of the presumed occultation event.

We provide in Table~\ref{tab:parameter} the relevant information of Bernhard-1 and Bernhard-2. The significance of the parallax measurement from the Gaia mission is at the level of 2--3$\sigma$, leading to unreliable distance inferences for both objects. Nevertheless, we find that Bernhard-1 may be associated with the Cygnus OB associations \citep{Quintana:2021}, whereas Bernhard-2 is not associated with any known star-forming groups or young star associations.

With the given coordinates we have also retrieved archival photometric observations. In particular, we found optical observations ($g$, $r$, $i$, $z$, and $y$) from the Panoramic Survey Telescope And Rapid Response System \citep[Pan-STARRS,][]{Chambers:2016, Flewelling:2020} between MJD=55,300--56,900, which provided useful constraints on the occultation model (see Section~\ref{sec:model}). Additionally, we found near-infrared observations from 2MASS ($J$, $H$, and $K_s$; \citealt{Skrutskie:2006, tmass_data}) and WISE ($W1$--4, \citealt{Wright:2010, wise_data}) taken at multiple epochs. These observations extend the spectral energy distribution (SED) and reveal the existence of the circumstellar (or circumbinary) disks (see Section~\ref{sec:model}).

\begin{deluxetable}{lcc}
\tablecaption{Candidate information.} \label{tab:parameter}
\tablehead{\colhead{} & \colhead{Bernhard-1} & \colhead{Bernhard-2}}
\startdata
ZTF identifier & J202055.22+381323.1 & J071445.39-090152.1 \\
RA$_{\rm J2000}$ & $20^{\rm h}20^{\rm m}55\fs22$ & $07^{\rm h} 14^{\rm m} 45\fs39$ \\
Decl$_{\rm J2000}$ & $+38^\circ13^\prime 23\farcs1$ & $-9^\circ 01^\prime 52\farcs1$ \\
Parallax$^a$ (mas) & $0.59\pm0.16$ & $0.34\pm0.16$ \\
$P$ (days)  & $192.10\pm0.02$ & $63.358\pm0.003$ \\
$t_{\rm in}$ (MJD) & $58227.81\pm0.07$ & $59155.24\pm0.04$ \\
$t_{\rm out}$ (MJD) & $58339.29\pm0.04$ & $59181.18\pm0.03$ \\
$v_{\rm in}$ ($R_\star$/day) & $0.134\pm0.003$ & $0.416\pm0.010$ \\
$v_{\rm out}$ ($R_\star$/day) & $0.150\pm0.002$ & $0.291\pm0.003$ \\
\enddata
\tablecomments{$^a$ Taken from Gaia Early Data Release 3 \citep{Gaia_EDR3}. These values are not changed in Gaia DR3.}
\end{deluxetable}

\section{Disk-occultation systems} \label{sec:model}

\begin{figure}
    \centering
    \includegraphics[width=0.9\columnwidth]{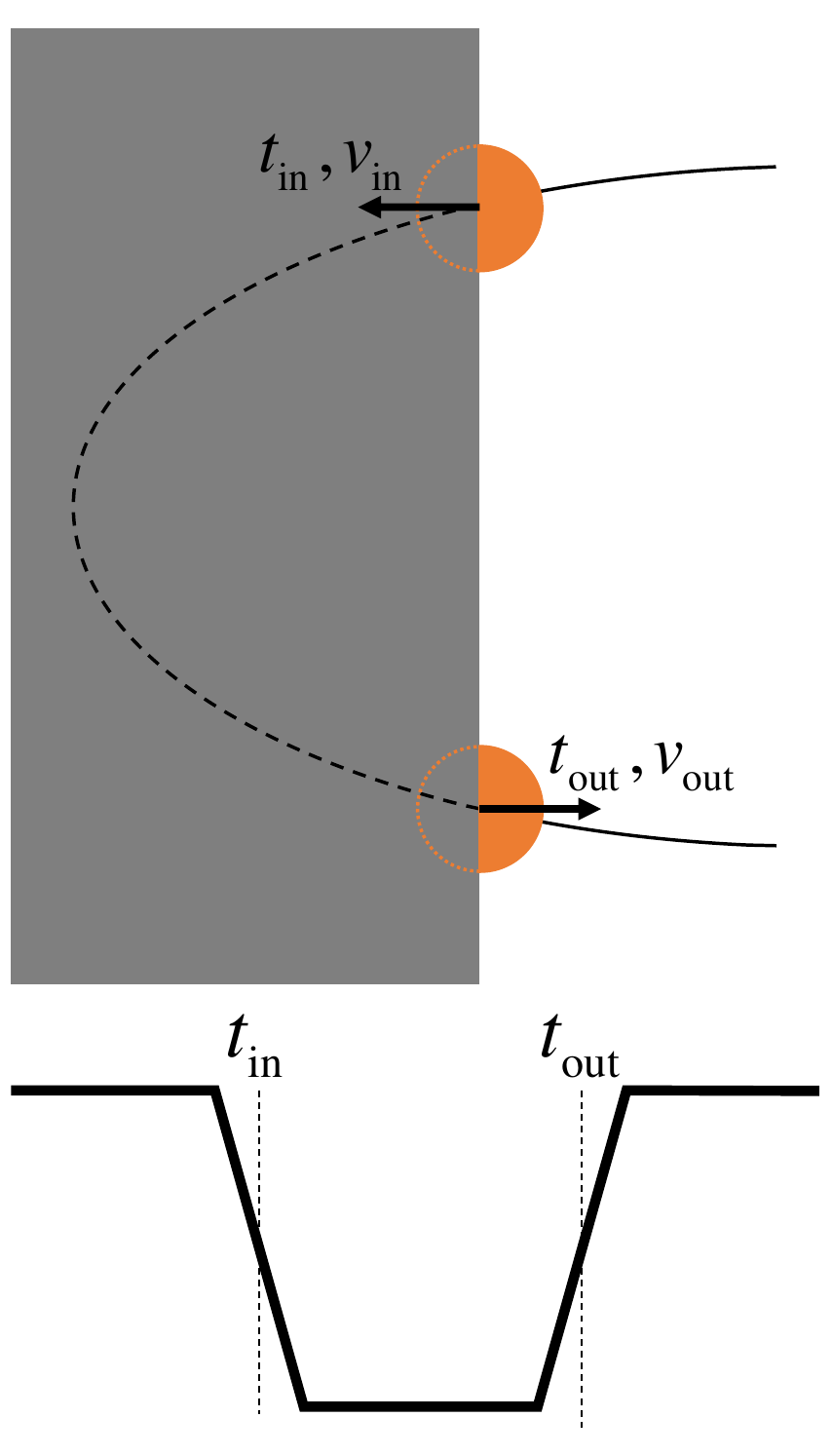}
    \caption{Schematic of the one-sided, fully opaque screen model used in this work. The model is described by five parameters: the time and perpendicular velocity of the star at epochs of ingress (with subscript ``in'') and egress (with subscript ``out''), respectively, and the orbital period of the star. The bottom panel shows an example light curve from the model.}
    \label{fig:model}
\end{figure}

\begin{figure*}
    \centering
    \includegraphics[width=\textwidth]{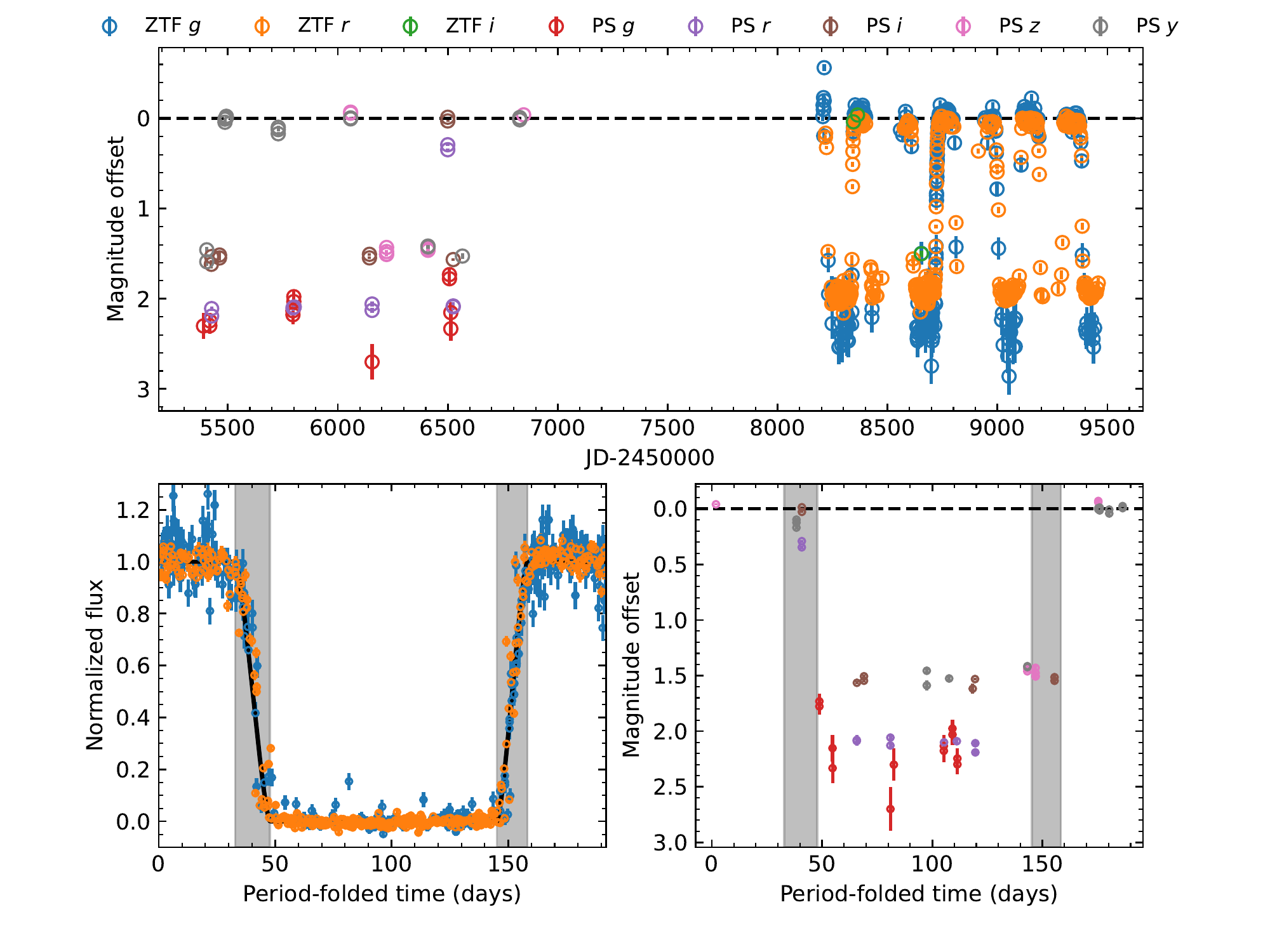}
    \caption{Bernhard-1 light curves. The upper panel shows the available data from ZTF survey and the archival Pan-STARRS (PS) data release 2. The out-of-occultation magnitudes have been subtracted to align the different bandpasses. The lower left panel shows the period-folded and flux-normalized ZTF $g$ and $r$ light curves on top of the model light curve. The lower right panel is the Pan-STARRS data folded by the same period. The shaded regions in the lower panels indicate the durations of ingress and egress. The best-fit model based on the ZTF data cannot match the Pan-STARRS data, suggesting the breakdown of the simplified model or the evolution of the occultation profile, potentially due to a precessing circumbinary disk.}
    \label{fig:bernhard1}
\end{figure*}

\begin{figure*}
    \centering
    \includegraphics[width=\textwidth]{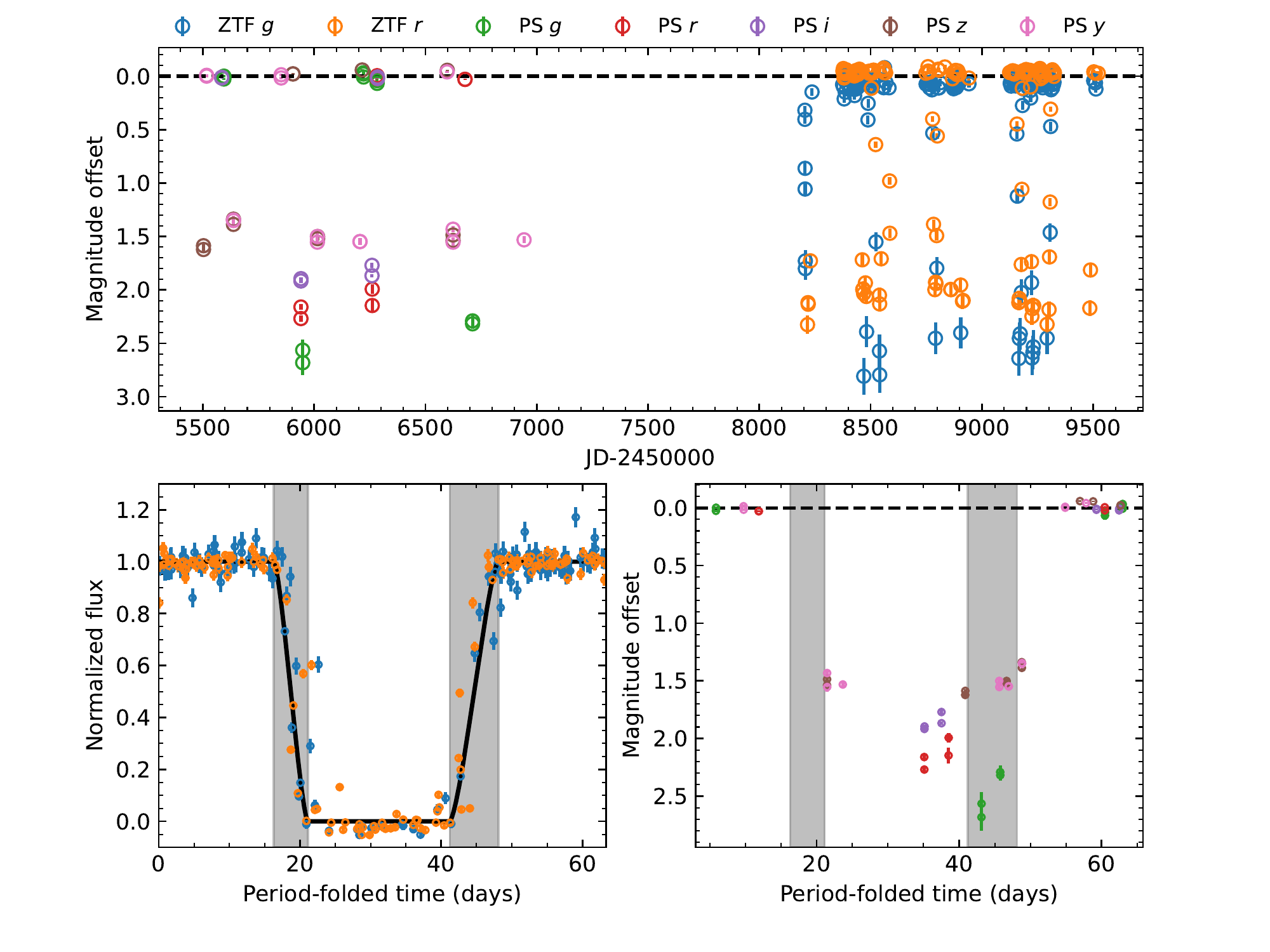}
    \caption{Similar to Figure~\ref{fig:bernhard1} but for Bernhard-2 system.}
    \label{fig:bernhard2}
\end{figure*}

\begin{figure*}
    \centering
    \includegraphics[width=0.45\textwidth]{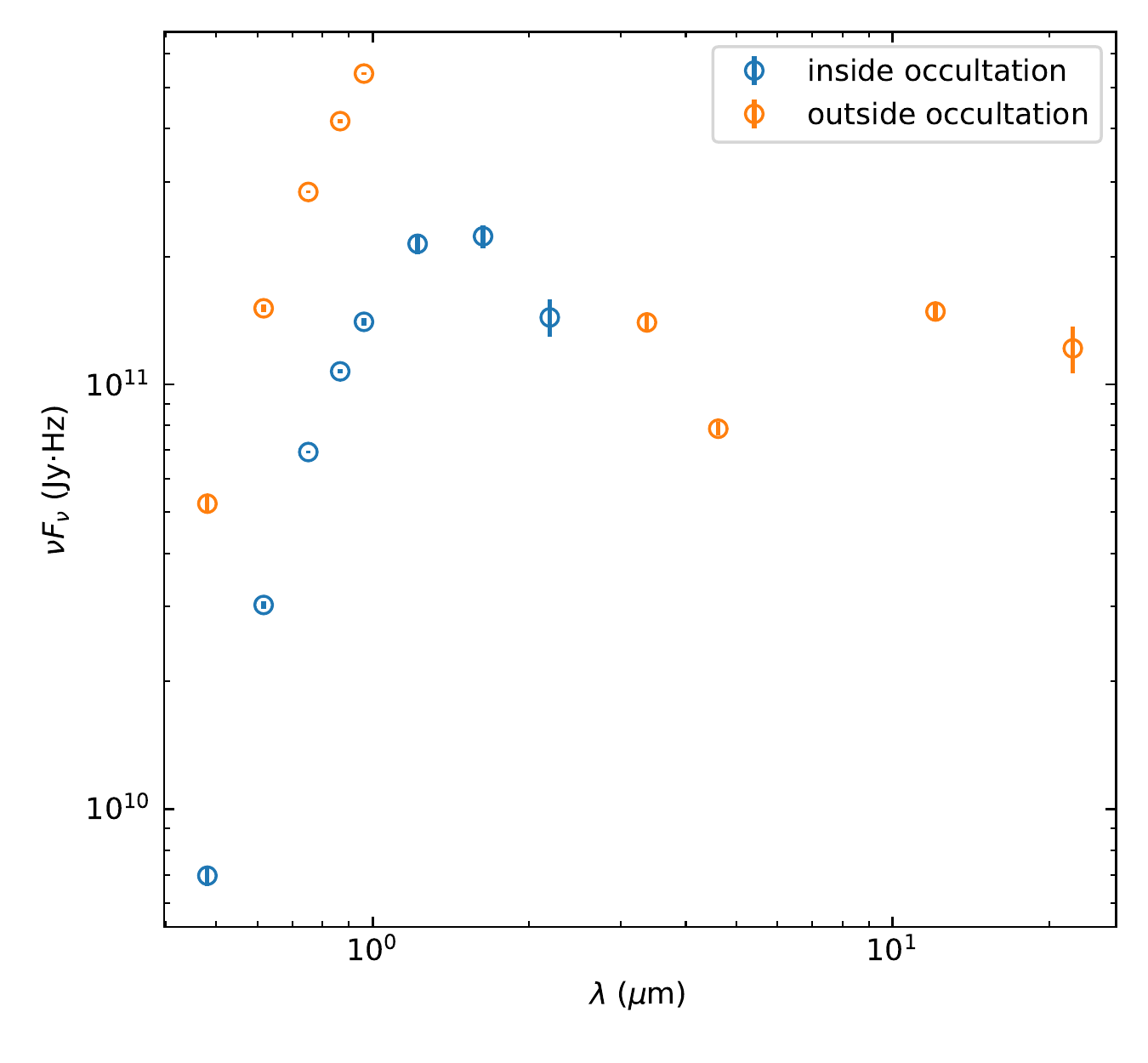}
    \includegraphics[width=0.45\textwidth]{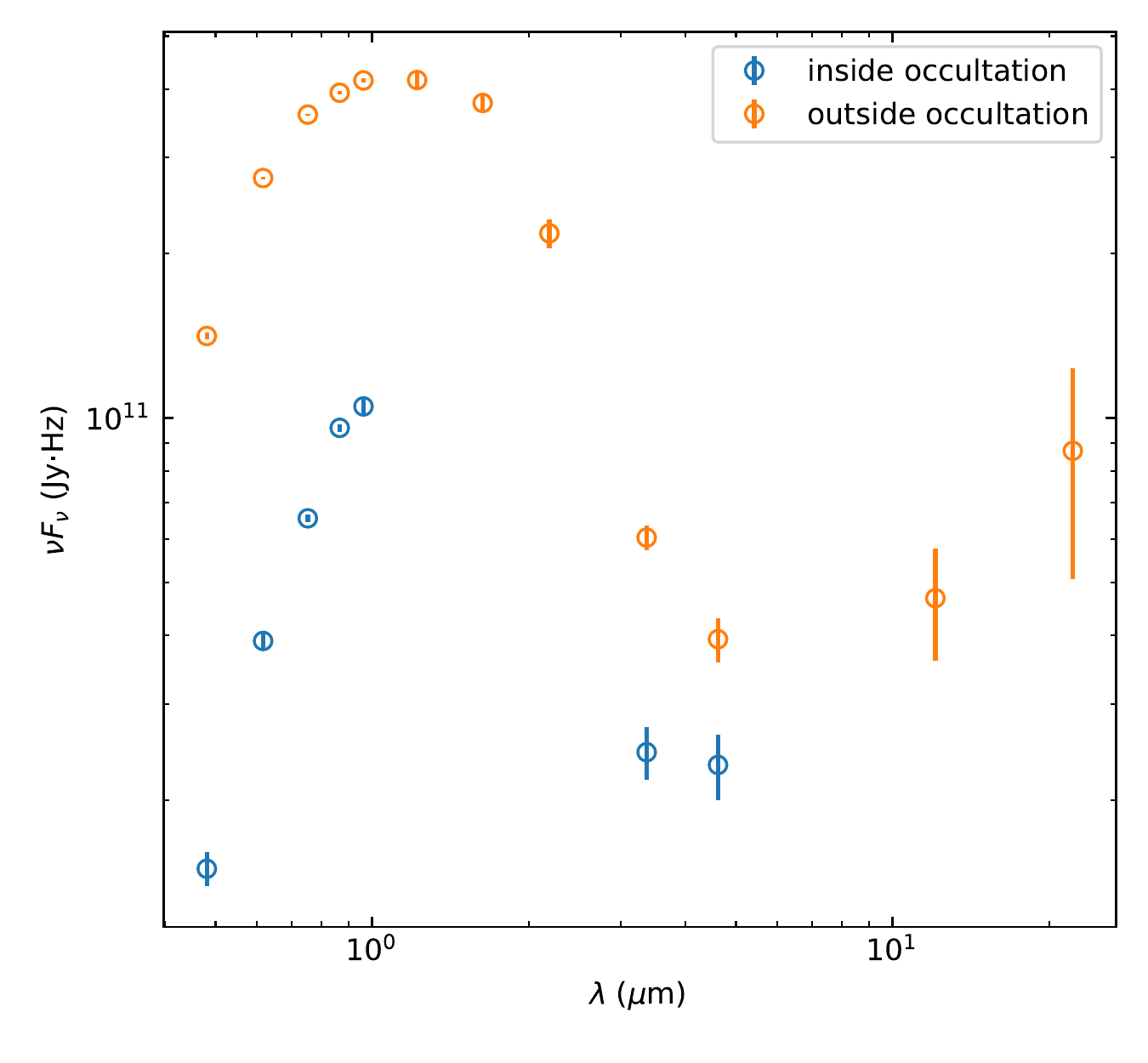}
    \caption{SEDs of both Bernhard objects, suggesting the presence of a cold disk component. The measurements used here are from Pan-STARRS ($g$, $r$, $i$, $z$, $y$), 2MASS ($J$, $H$, $K_s$), and WISE ($W1$--4). SED measurements are also available in Table~\ref{tab:sed} in terms of magnitude values.}
    \label{fig:sed}
\end{figure*}

\begin{deluxetable}{lcccc}
\tablecaption{SED information of both candidate objects. For each object, magnitudes inside and outside the occultation are given.} \label{tab:sed}
\tablehead{\multirow{2}{*}{Filter} & \multicolumn{2}{c}{Bernhard-1} & \multicolumn{2}{c}{Bernhard-2} \\
& \colhead{in} & \colhead{out} & \colhead{in} & \colhead{out}}
\startdata
$g$ & $21.28(6)$ & $19.09(6)$ & $20.46(8)$ & $18.01(2)$ \\
$r$ & $19.42(2)$ & $17.30(2)$ & $19.14(4)$ & $17.02(1)$ \\
$i$ & $18.301(7)$ & $16.770(7)$ & $18.36(2)$ & $16.514(4)$ \\
$z$ & $17.67(1)$ & $16.20(1)$ & $17.80(2)$ & $16.261(6)$ \\
$y$ & $17.27(2)$ & $15.806(6)$ & $17.658(3)$ & $16.09(1)$ \\
$J$ & $15.64(6)$ & \nodata & \nodata & $14.92(4)$ \\
$H$ & $14.80(7)$ & \nodata & \nodata & $14.27(4)$ \\
$K_s$ & $14.5(1)$ & \nodata & \nodata & $14.07(7)$ \\
$W_1$ & \nodata & $13.21(5)$ & $15.10(12)$ & $14.15(6)$ \\
$W_2$ & \nodata & $12.85(4)$ & $14.21(15)$ & $13.60(10)$ \\
$W_3$ & \nodata & $9.29(5)$ & \nodata & $10.56(25)$ \\
$W_4$ & \nodata & $7.4(1)$ & \nodata & $7.8(5)$ \\
\enddata
\end{deluxetable}

\subsection{Occultation model}

In both candidate systems, the occultations last for $\sim50\%$ of the total period. This cannot be explained by a circumstellar disk occulting a binary star at an exterior orbit, as seen in EE Cephei, $\epsilon$ Aurigae and a few other similar systems \citep[e.g.,][]{Mikolajewski:1999, Dong:2014, Zhou:2018}. Additionally, the occultation periods, $192$ and $63$ days, are too short to be explained by a precessing disk occulting one single central object. Therefore, we conclude that the circumbinary disk occulting a central binary is the most plausible explanation for both systems. They are therefore KH 15D-like.

Given the scarce photometric observations, we do not apply the detailed precessing disk model, as was developed for the case of KH 15D \citep[e.g.,][]{Chiang2004, Winn:2006, Poon:2021}. Instead, a simplified one-sided, fully opaque screen model, as illustrated in Figure~\ref{fig:model}, is used to describe the time evolution of the occultation event. This model involves five primary parameters: $t_{\rm in}$ ($t_{\rm out}$) and $v_{\rm in}$ ($v_{\rm out}$) are the epoch and projected perpendicular velocity of the star at the middle of ingress (egress), respectively, and $P$ is the period of the occultation (i.e., the inner binary).

Specifically, we model the stellar brightness profile as
\begin{equation}
    F(t) = F_1 \cdot f(t) + F_2 .
\end{equation}
Here $F_1+F_2$ and $F_2$ are the flux values of the system outside and inside the occultation, respectively. These linear parameters are derived analytically for any given set of model parameters $(t_{\rm in}, t_{\rm out}, v_{\rm in}, v_{\rm out}, P)$ through the maximum likelihood method. The normalized light curve $f(t)$ is given by
\begin{equation}
    f(t) = \left\{ \begin{array}{ll}
        1 &  , x \leq -1 \\
        \frac{1}{2} - \frac{1}{\pi} \left[ x\sqrt{1-x^2} + \arcsin{x} \right] & , -1<x<1 \\
        0 & , x \geq 1 
    \end{array} \right. ,
\end{equation}
where 
\begin{equation}
    x = \left\{ \begin{array}{ll}
        v_{\rm in} (t-t_{\rm in}) & ,{\rm ingress} \\
        v_{\rm out} (t_{\rm out}-t) & ,{\rm egress}
    \end{array} \right. .
\end{equation}
Here we have assumed no limb-darkening effect for the stellar surface.

This one-sided screen model is applied to the ZTF data of both Bernhard-1 and Bernhard-2 objects. The \texttt{emcee} sampler from \citet{ForemanMackey:2013} is used to perform the Markov chain Monte Carlo (MCMC) analysis and derive the uncertainties on the model parameters. The results from this analysis are given in Table~\ref{tab:parameter}. We then apply the best-fit models to the archival Pan-STARRS data, which went back as long as $\sim11$ years. For each target, we show the joint photometric light curves from ZTF and Pan-STARRS as well as the period-folded light curves of individual data sets. These are illustrated in Figures~\ref{fig:bernhard1} and~\ref{fig:bernhard2}. The same models are also used to construct the SEDs when the targets are inside and outside of the disk occultation. The resulting SEDs are shown in Figure~\ref{fig:sed}.

\subsection{Bernhard-1}

As shown in Figure~\ref{fig:bernhard1}, the occultation event in Bernhard-1 has a period of 192 days with a duration of $112\,$days, or $58\%$ of the total binary period. The stellar brightness remains fairly constant outside of the occultation, with no signature of active accretion. The available observations are also too sparse to allow any detection of the stellar rotation. Inside the occultation, the light curve is also very flat. In particular, there is not re-brightening in the middle of the occultation due to the emergence of the companion star, as was seen in the case of KH 15D. If the central object is indeed a binary, one possible explanation is then that the binary has fairly large eccentricity such that the companion star does not emerge when the primary is deepest into the occulting disk. This is supported by the long duration (i.e., $>50\%$ of the binary period) of the occultation and the asymmetric ingress/egress features. Additionally, eccentric binaries can help keep the surrounding circumbinary disk misaligned \citep[e.g.,][]{Martin:2017, Zanazzi:2018, Smallwood:2019}. Alternatively, the companion star may be very faint or even dark. Future spectroscopic observations may help resolve this issue.

With a binary period of $P=192\,$d, the separation between the binary components should be $a=0.65\,{\rm au} (M_{\rm tot}/M_\odot)^{1/3}$, where $M_{\rm tot}$ is the combined mass of the binary. Since the circumbinary disk is likely tidally truncated \citep[e.g.,][]{Miranda:2015, Lubow:2015}, its inner radius is likely $\gtrsim$au. Better knowledge of the binary as well as the disk properties is needed in order to determine the exact value of the inner boundary.

The best-fit model, derived from the ZTF data, does not seem to match the archival data from Pan-STARRS, as shown in the lower right panel of Figure~\ref{fig:bernhard1}. This suggests the evolution of the system light curves, potentially due to the precession of the circumbinary disk \citep[e.g.,][]{Chiang2004, Winn:2004, Poon:2021}. It may also suggest the failure of our overly-simplified model. Long-term photometric observations are encouraged to place tighter constraints on the light curve evolution.

As shown in the left panel of Figure~\ref{fig:sed}, the near-infrared excess in the SED of Bernhard-1 suggests the existence of a cold disk component. This is consistent with the expectation that a stellar binary is being occulted by a tilted circumbinary disk.

\subsection{Bernhard-2}

\begin{figure}
    \centering
    \includegraphics[width=\columnwidth]{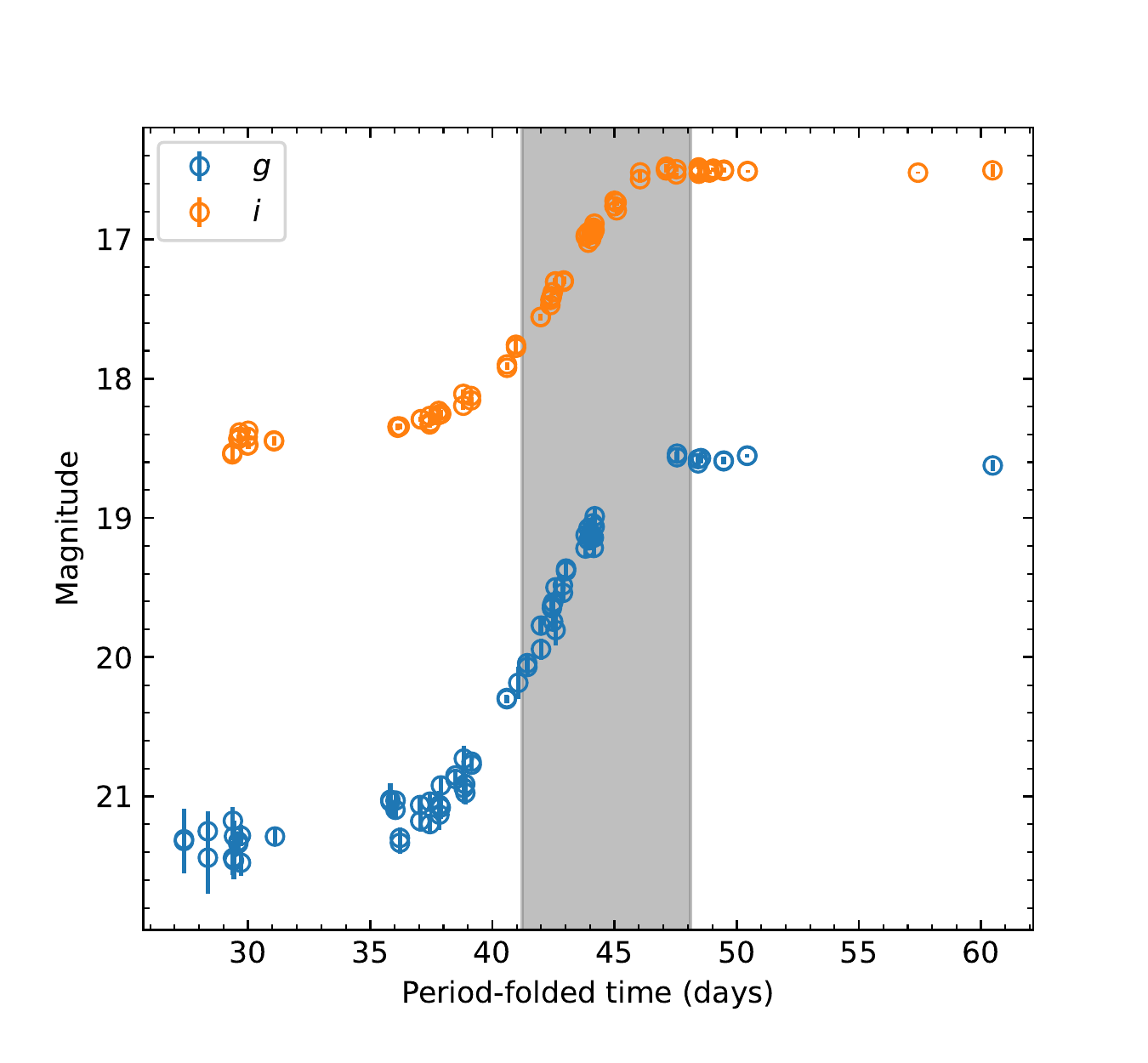}
    \caption{LCOGT observations in $g$ and $i$ bands on Bernhard-2 object. The light curves are folded in the same way as the ZTF data in the lower left panel of Figure~\ref{fig:bernhard2}. The shaded region marks the egress according to the one-sided screen model.}
    \label{fig:lco-data}
\end{figure}

Apart from a shorter period (63 days), Bernhard-2 shares several similar features with Bernhard-1: a flat light curve outside of occultation, asymmetric ingress/egress regions, and the lack of re-brightening at the middle of occultation. The ingress and egress regions in the period-folded ZTF light curve have large scatterings. Together with the fact that the best-fit model cannot explain the archival Pan-STARRS observations, it again may suggest the precession of the disk and/or the failure of the one-sided screen model.

We attempted to obtain follow-up, high-cadence photometric observations of Bernhard-2 when it was visible from the ground. A few nights of observations were obtained from the 32-inch telescope at the Post Observatory, the 16-inch telescope at the Remote Observatory Atacama Desert (ROAD, \citealt{Hambsch:2012}), and the 1-m telescopes of the Las Cumbres Observatory Global Telescope \citep[LCOGT,][]{Brown:2013} in December of 2021, when the system was exiting the egress. In early 2022, more systematic observations were obtained on the 1-m LCOGT telescopes and captured a broader range of the egress region. The LCOGT observations are reduced by \texttt{AstroImageJ} \citep{Collins:2017} and shown in Figure~\ref{fig:lco-data}, which indicate a rather smooth and gradual transition from inside to outside of the occultation. Our follow-up photometric observations are available via an online-only table.

We observed Bernhard-2 with the High Resolution Echelle Spectrometer on the 10m Keck I telescope (Keck/HIRES) at UT 07:42 on Jan 8, 2022, when the object was outside the occultation. At a seeing of $1.4\arcsec$, we integrated for 20\,min without the iodine cell and adopted the master wavelength solution for that night. We achieved a S/N of $\sim18$ per resolution limit. To retrieve the spectroscopic parameters, we selected five spectrum segments in ranges of 5350--5500, 5500--5750, 5910--6180, 6320--6410, and 6670--6780 \AA, respectively, and performed matches to the ELODIE library \citep{elodie}. This leads to a surface effective temperature $T_{\rm eff}=4865\pm82\,$K and a surface gravity $\log{g}=4.37\pm 0.04$, and the object is found to be a K$1.3\pm0.5$ type pre-main-sequence star \citep{Pecaut:2013}. This spectroscopic classfication is consistent with the broadband colors of Bernhard-2 outside the occultation (see Table~\ref{tab:sed}), assuming a small amount of extinction \citep{Fang:2017}.
Unlike the spectrum of KH 15D \citep[e.g.,][]{Fang:2019}, the spectrum of Bernhard-2 contains no emission lines. This confirms the result from photometric observations that the object has no accretion activity and thus is probably an older star than KH 15D.

The WISE $W3$- and $W4$-band observations had relatively low S/N values, as shown in the right panel of Figure~\ref{fig:sed}. Nevertheless, the combined SED indicates the existence of a cold disk component, which is consistent with the expectation that the system is surrounded by a circumbinary disk.

\section{Discussion}

This paper presents the discovery and analysis of two systems that show disk occultation signatures. These systems, named Bernhard-1 and Bernhard-2, show periodic dimmings that have large amplitudes ($>1.5$ mag in both $g$ and $r$) and relatively long durations ($\gtrsim50\%$ of the identified periods of 192 and 63 days, respectively). Both inside and outside of the occultations, the light curves appear to be fairly flat with no clear signs of active accretion. These features are best explained by the tilted disk occulting a central binary, similar to the famous case of KH 15D \citep[e.g.,][]{Poon:2021}. SEDs of both systems indeed suggest the existence of cold disk components.

A one-sided, fully opaque screen model is used to fit the photometric observations from ZTF and Pan-STARRS, which have a time span of up to $\sim$11\,yr. Although this model cannot explain the details of the photometric light curve, especially the portions of ingress and egress, it can reasonably reproduce the key features of the observed data. According to this model, the circumbinary disks in both systems may have been gradually precessing on a timescale of $\sim$10\,yr. The current observations are too sparse, and the orbital properties of the central binaries remain unknown, preventing us from applying a more complicated occultation model. Both photometric and spectroscopic observations are encouraged in order to reveal the true nature of these systems.

Our findings represent a valuable addition to the rare class of KH 15D-like systems. Prior to this work, the only known objects in this class other than KH 15D are WL4 and YLW 16A in the $\rho$ Oph star forming region, both of which were identified in the near infrared and highly extincted in the optical \citep{Plavchan:2008, Plavchan:2013}. Taking the number of candidate young stellar objects from the data release 3 of Gaia \citep{Marton:2022}, which is comparable to ZTF in terms of the magnitude limit, we find $\sim$3/80,000 for the rate of KH 15D-like objects in the optical survey down to $\sim20\,$mag. This very rough estimate only applies to stellar binaries with relative long ($>10\,$d) periods, and it is possible that young binaries with shorter periods may also show disk occultation signatures. We leave the more complete search for such systems to future works.

\begin{acknowledgments}
We thank Richard Post for contributing to the photometric observations of Bernhard-2 and Greg Herczeg for useful discussion.
We also thank the anonymous referee for comments and suggestions on the manuscript.
W.Zhu, W.Zang, and T.G.\ acknowledge support from the National Natural Science Foundation of China (grant No. 12133005). 
W.Zhu and S.D.\ are supported by the science research grant from the China Manned Space Project with No.\ CMS-CSST-2021-B12.
W.Zhu is also supported by the National Science Foundation of China (grant No.\ 12173021). 
J.Z. was supported by the Natural Sciences and Engineering Research Council of Canada (NSERC) under the funding reference \# CITA 490888-16.
This research uses data obtained through the Telescope Access Program (TAP), which has been funded by the TAP member institutes.
Based on observations obtained with the Samuel Oschin Telescope 48-inch and the 60-inch Telescope at the Palomar Observatory as part of the Zwicky Transient Facility project. ZTF is supported by the National Science Foundation under Grants No. AST-1440341 and AST-2034437 and a collaboration including current partners Caltech, IPAC, the Weizmann Institute for Science, the Oskar Klein Center at Stockholm University, the University of Maryland, Deutsches Elektronen-Synchrotron and Humboldt University, the TANGO Consortium of Taiwan, the University of Wisconsin at Milwaukee, Trinity College Dublin, Lawrence Livermore National Laboratories, IN2P3, University of Warwick, Ruhr University Bochum, Northwestern University and former partners the University of Washington, Los Alamos National Laboratories, and Lawrence Berkeley National Laboratories. Operations are conducted by COO, IPAC, and UW.
This publication makes use of data products from the Two Micron All Sky Survey, which is a joint project of the University of Massachusetts and the Infrared Processing and Analysis Center/California Institute of Technology, funded by the National Aeronautics and Space Administration and the National Science Foundation.
This publication makes use of data products from the Wide-field Infrared Survey Explorer, which is a joint project of the University of California, Los Angeles, and the Jet Propulsion Laboratory/California Institute of Technology, funded by the National Aeronautics and Space Administration.
Some of the data presented herein were obtained at the W.\ M.\ Keck Observatory, which is operated as a scientific partnership among the California Institute of Technology, the University of California and the National Aeronautics and Space Administration. The Observatory was made possible by the generous financial support of the W. M. Keck Foundation. 
\end{acknowledgments}

\bibliography{my_bib}{}
\bibliographystyle{aasjournal}



\end{CJK*}
\end{document}